\documentclass[a4paper,11pt]{article}
\usepackage{pos}
\usepackage{epsfig}

\title{How neutrinos could help solving cosmological anomalies and tensions}

\author{Pasquale Di Bari}

\affiliation{School of Physics and Astronomy, University of Southampon\\
 Southampton, SO17 1BJ, U.K.}

\emailAdd{P.Di-Bari@soton.ac.uk}

\abstract{In this talk I discuss how neutrinos might help solving or alleviating different anomalies and tensions in cosmology.
Invisible decays of the heaviest relic neutrinos might provide a way to solve the 
neutrino mass tension between cosmological observations and neutrino oscillation experiments.
The excess radio background mystery could be explained by radiative decays of relic neutrinos.
However, the upper bound on the neutrino effective magnetic moment requires some trick to be circumvented.
To this extent, I discuss a recently proposed boomerang mechanism in which the visible sector throws dark neutrinos into the dark sector at $t \sim 100\,{\rm s}$
and $T \sim 100\,{\rm keV}$, and much later (basically at the present time) the dark sector throws back photons into the visible sector.  
The mechanism predicts an effective neutrino magnetic moment that might be within the reach of next experiments.
Some contribution to the 21 cm cosmological signal is also expected.
These are exciting times for cosmological searches of BSM physics. 
}

\FullConference{Proceedings of the Corfu Summer Institute 2025 "School and Workshops on Elementary Particle Physics and Gravity" (CORFU2025)\\
27 April - 28 September, 2025\\
Corfu, Greece\\
This talk is mainly based on \cite{Dev:2023wel,Dev:2025ufo}.
Slides of the talk can be found at: https://www.southampton.ac.uk/~pdb1d08/PDB4CORFU25CT.pdf}

\begin{document}

\def\mc#1{\mathcal#1}
\def\a{\alpha}
\def\b{\beta}
\def\c{\chi}
\def\d{\delta}
\def\e{\epsilon}
\def\f{\phi}
\def\g{\gamma}
\def\h{\eta}
\def\i{\iota}
\def\j{\psi}
\def\k{\kappa}
\def\la{\lambda}
\def\m{\mu}
\def\n{\nu}
\def\o{\omega}
\def\p{\pi}
\def\q{\theta}
\def\r{\rho}
\def\s{\sigma}
\def\t{\tau}
\def\u{\upsilon}
\def\x{\xi}
\def\z{\zeta}
\def\D{\Delta}
\def\F{\Phi}
\def\G{\Gamma}
\def\J{\Psi}
\def\L{\Lambda}
\def\O{\Omega}
\def\P{\Pi}
\def\Q{\Theta}
\def\S{\Sigma}
\def\U{\Upsilon}
\def\X{\Xi}

\def\ve{\varepsilon}
\def\vf{\varphi}
\def\vr{\varrho}
\def\vs{\varsigma}
\def\vq{\vartheta}

\newcommand{\vev}[1]{\langle #1 \rangle}
\def\dg{\dagger}                                     
\def\ddg{\ddagger}                                   
\def\wt#1{\widetilde{#1}}                    
\def\mt{\widetilde{m}_1}
\def\mti{\widetilde{m}_i}
\def\mtj{\widetilde{m}_j}
\def\rt{\widetilde{r}_1}
\def\mtt{\widetilde{m}_2}
\def\mttt{\widetilde{m}_3}
\def\rtt{\widetilde{r}_2}
\def\mb{\overline{m}}
\def\VEV#1{\left\langle #1\right\rangle}        
\def\be{\begin{equation}}
\def\ee{\end{equation}}
\def\ds{\displaystyle}
\def\ra{\rightarrow}

\def\bea{\begin{eqnarray}}
\def\eea{\end{eqnarray}}

\maketitle

\section{Introduction}

 Extensions of the standard model are very strenuously and effectively tested by colliders that are so far placing more and more stringent constraints,
pushing the border where new physics could lie at higher energy scales and smaller couplings. However, new physics could lie
in a {\em deep unknown} region. We know that the existence of new physics is very strongly motivated by the necessity of addressing the 
cosmological puzzles and explaining neutrino masses and mixing. This could sound as a frustrating situation. 
 Fortunately, in the last years, a host of new ideas, tools and anomalies in cosmology and astrophysics have hugely expanded our ways to
explore this deep unknown region, circumventing the intrinsic limitations  of collider machines. 

The map in Fig.~1 shows the variety of directions one can explore nowadays. Hopefully, somewhere, we will finally discover the new physics
able to address the puzzle of the origin of matter in the universe (dark matter and matter-antimatter asymmetry) and the origin of neutrino masses.
\begin{figure}[t]
\hspace{25mm}\psfig{file=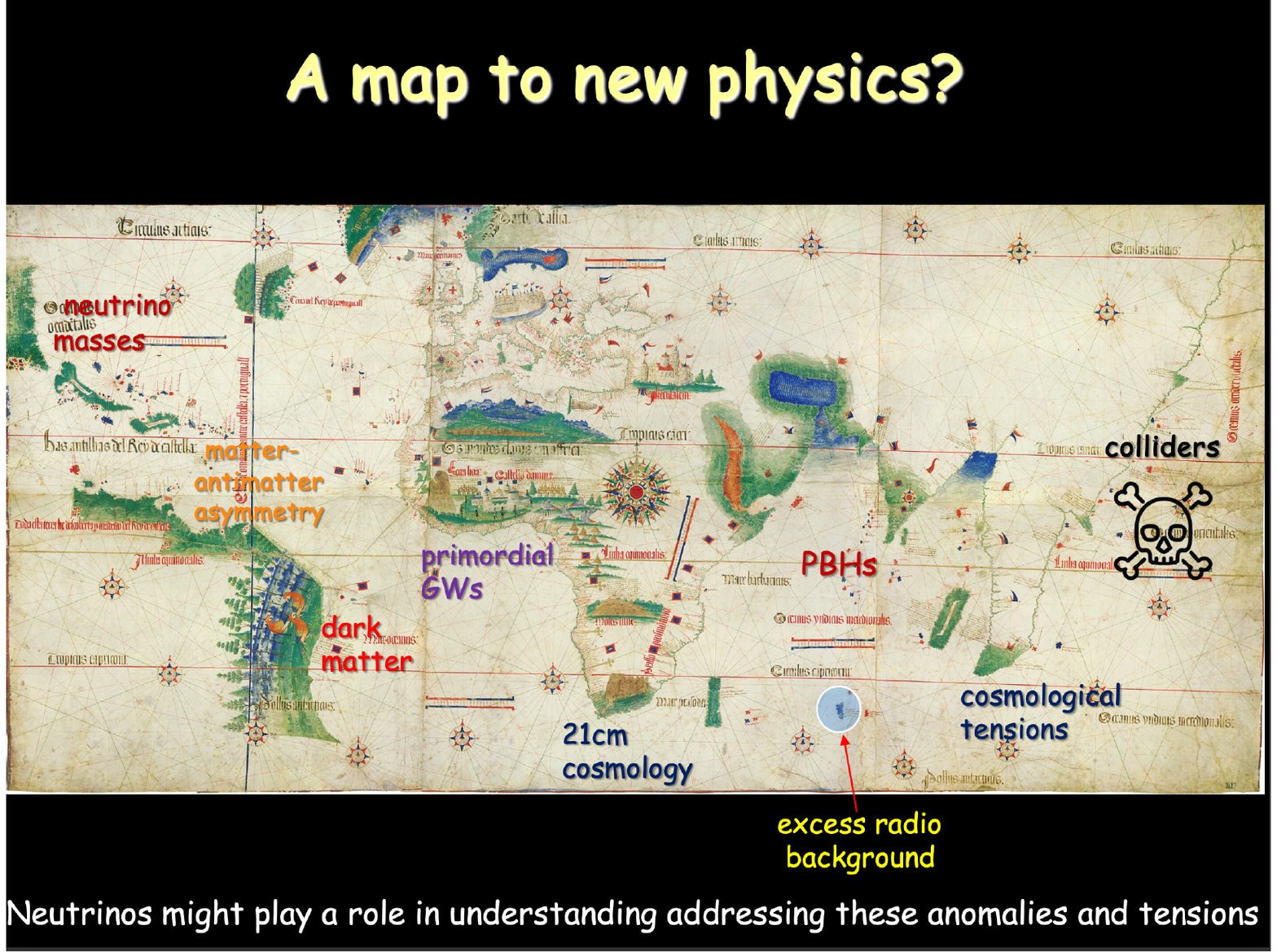,scale=0.35,angle=-90} \caption{Map of new physics hints, evidences, opportunities.
The background map is the famous {\em Cantino planisphere} (1502).}
\end{figure}
As stated in the figure, neutrinos might play an important role in addressing some of the cosmological anomalies and tensions. 

\section{Neutrino mass tension}

Neutrino oscillation experiments place a lower bound on the sum of neutrino masses \cite{Esteban:2024eli}
\be\label{lowerb}
\sum_i m_{\nu_i} \geq  58 \,{\rm meV} \,  \hspace{10mm}  \mbox{\rm (95 \% C.L.)} \,  .
\ee
From cosmological observation the {\em Planck} collaboration has placed the upper bound \cite{Tristram:2023haj}
\be\label{upperb}
\sum_i m_i \leq 0.11 \, {\rm eV} \hspace{10mm}  \mbox{\rm (95 \% C.L.)} \,  ,
\ee
combining data from CMB temperature and polarisation anisotropies, CMB lensing and baryon acoustic oscillations (BAO) data  from 6dF and SDSS galaxy surveys.
The best fit value is simply $\sum_i m_i = 0$, implying that cosmology is currently not sensitive to neutrino masses.  It also implies that
there is a tension with the lower bound (\ref{lowerb}) from neutrino oscillation experiments, at the level of $\sim 1.5 \s$. 
This tension is even more exacerbated by a  combination of {\em Planck} CMB data with the recent BAO data from the DESI DR2 galaxy survey  \cite{DESI:2025zgx}, finding:
\be
\sum_i m_i \leq 65 \, {\rm meV} \hspace{10mm}  \mbox{\rm (95 \% C.L.)} \,  .
\ee
In this case the tension is almost at $2\sigma$. It should be stressed that the above upper bounds from cosmology 
assume the standard $\Lambda$CDM model.  Many different ways to remove this tension, relaxing the upper bounds, have been proposed by
extending the $\Lambda$CDM model. For example, the same DESI collaboration finds the upper bound \cite{DESI:2025zgx}
\be
\sum_i m_i \leq 160 \, {\rm meV} \hspace{10mm}  \mbox{\rm (95 \% C.L.)} \,  ,
\ee
within a $w_0 w_a$ model where the dark energy equation of state parameter $w(a)$ is not constant but 
evolves with the scale factor $a$ as
\be
w(a) = w_0 + w_a (1-a) \,  ,
\ee
with $w_0=-0.42^{+0.24}_{-0.21}$ and $w_a = -1.75\pm 0.63$.  As a second example, within a model assuming
a suppressed matter perturbation growth rate, the authors of \cite{Giare:2025ath} find
\be
\sum_i m_i \leq 134 \, {\rm meV} \hspace{10mm}  \mbox{\rm (95 \% C.L.)} \,  .
\ee 
However, there is also a simple particle physics solution, not requiring an extension of the $\Lambda$CDM model.\footnote{One could think that a solution in terms of  an extension of the $\Lambda$CDM model is favoured  also by the existence of the other cosmological tensions, such as the Hubble tension. Unfortunately, solutions solving the neutrino mass tension do not seem to help solving the other cosmological tensions.}

Relic neutrinos could decay with a lifetime lower than (at least) one order of magnitude the age of the universe but longer than
$\sim 10^9\,{\rm s}\,(\sum_i m_i/50 \, {\rm meV})^3$ not to clash with CMB anisotropy observations (free streaming neutrinos need to be present 
during recombination) \cite{Hannestad:2005ex,Serpico:2007pt,Archidiacono:2013dua,Escudero:2019gfk,Escudero:2020ped,Craig:2024tky}. 
With such short lifetimes, radiative neutrino decays are strongly constrained by the upper bound on CMB spectral distortions from the FIRAS instrument on the
COBE satellite. Therefore, they necessarily need to decay invisibly.  

For example, if active neutrinos interact with a scalar field $\phi$ with interactions \cite{Escudero:2020ped,Craig:2024tky}
\be
{\cal L}_{\nu-\phi} = {\lambda_{ij} \over 2}\,\bar{\nu}_i \nu_j \phi  +{\rm h.c.} \,   ,
\ee 
one would open the decay channel $\nu_i \rightarrow \nu_j + \phi$ with lifetime
\be
\tau_{\nu_i\rightarrow\nu_j +\phi} \simeq 7 \times 10^{17}\,{\rm s} \left({0.05\,{\rm eV} \over m_{\nu_i}}\right)
 \, \left({10^{-15} \over \lambda_{ij}^2}\right)^2 \,   .
\ee
Therefore, the tension between oscillation neutrino experiments and cosmological observations 
might be interpreted as if it suggests the existence of a low scale dark sector destabilising the 
cosmic neutrino background.  

\section{The excess radio background mystery}

The FIRAS instrument of COBE has measured the spectrum of the cosmic microwave background (CMB) in the frequency range (60 -- 600) GHz, corresponding approximately to the photon energy range ($2.5 \times 10^{-4}$--$2.5\times 10^{-3}$)\,{\rm eV}. 
In this frequency range, it has placed very strong upper bounds on deviations from a Planckian spectrum 
with temperature $T_{\gamma 0} = (2.7255 \pm 0.0006)\,{\rm K}$ \cite{Fixsen:2009ug}.  These translate into strong constraints in the parameter
of any model that predicts some amount of non-thermal radiation in that frequency range today.
However, the FIRAS instrument does not place constraints at frequencies below 60 GHz. 

The ARCADE 2 balloon-borne experiment has measured the absolute temperature of sky at frequencies in the range (3--10)\,GHz \cite{Fixsen:2009xn},
well below the FIRAS low threshold, covering a $8.4\%$ portion of the sky.  Only 6 data points gave in the end 
a meaningful result not dominated by noise. These point clearly show an excess compared to the CMB  temperature that is not
described by a thermal component (that implies that the temperature changes with the frequency).  
If we define $T_{\rm ERB}(\nu) \equiv T_{\gamma 0}(\nu) - T_{\rm CMB}$, where $T_{\gamma 0}(\nu)$ is the radiometric absolute 
temperature of the sky, the values of $T_{\rm ERB}(\nu)$ for the six data points are show in Fig.~2. 
\begin{figure}[t]
\begin{center}
\psfig{file=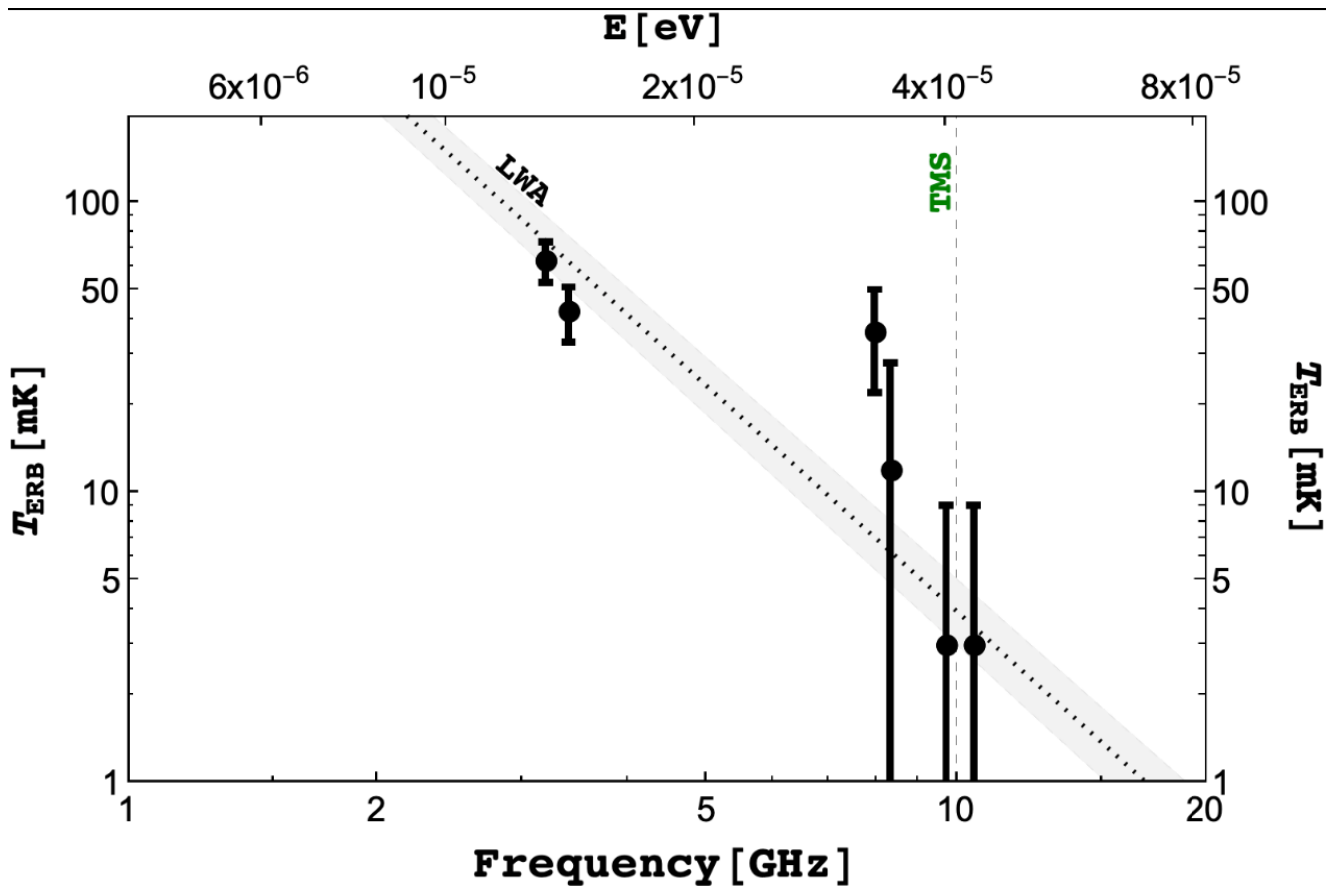,scale=0.35,angle=-90} \caption{Plot of $T_{\rm ERB}(\nu)$ measured
by ARCADE 2 for six values of $\nu$ as indicated. The grey  band is a power law fit, the dashed vertical line is the
low frequency threshold of the Tenerife Microwave Spectrometer (courtesy of Rishav Roshan).}
\end{center}
\end{figure}
When statistical errors are taken into account, the statistical significance of the excess is $\sim 5\sigma$. 
The excess cannot be explained in terms of a known population of radio sources and different attempts have not detected any anisotropy so that the source of the
excess has to be extremely smooth and this seems to suggest that can be better described by a smooth background rather than unknown radio sources. 
It also disfavours solutions where the signal is somehow correlated with the dark matter distribution, since one would expect some level of anisotropy
that is excluded by the observations. It is quite fair to conclude that the `nature of the background is still unknown' 
\cite{Grebenev:2024nkw}, 
and for this reason the excess radio background should be regarded as a mystery. 

An intriguing possibility is that the source of non-thermal radiation that would explain the excess 
is produced by a neutrino species decaying radiatively \cite{Chianese:2018luo}:  
$\nu_i \ra \nu_{\rm f} + \gamma$, where $\nu_f$ is 
neutrino species in the final state and $\nu_i$ is one of the active neutrino species. 
Active-to-active neutrino decays cannot realise the necessarily small required mass splitting:
$\Delta m_i \equiv m_i - m_{\rm f}\ll 2.4 \times 10^{-4}\,{\rm eV} \simeq 60 \, {\rm GHz}$.
The reason is that such a small mass splitting, when neutrino oscillation experimental information is taken into account,
would necessarily imply quasi-degenerate neutrino masses $m_1 \simeq m_2 \simeq m_3 \simeq 1\,{\rm eV}$ that are 
excluded by the cosmological upper bound Eq.~(\ref{upperb}).
For this reason $\nu_{\rm f}$ has to be necessarily a new sterile neutrino state. 
Assume that the decaying active neutrino $\nu_i$ and the
new sterile neutrino state $\nu_{\rm s}$ are quasi-degenerate, so that $\Delta m_i \ll m_i$.  Moreover, assume that the active neutrinos
decay non-relativistically. This implies a lower bound on the neutrino mass $m_i \gg 1\,{\rm meV}$. 

With these assumptions, the photon produced by the decays are monochromatic at the time of decay $t_{\rm D}$
with energy simply given by $E_{\gamma}(z_{\rm D}) = \Delta m_i$, where $z_{\rm D}$ is the cosmological redshift at
the time of decay.  Taking into account the cosmological expansion, at the present time one has
\be\label{Eg0}
E_{0} = {\Delta m_i \over 1 + z_{\rm D}} \leq \Delta m_i \,   .
\ee
For definiteness, we will assume in the following that the decaying neutrino is the lightest one so that we have $\nu_1 \ra \nu_{\rm s} + \gamma$.

The energy density of non-thermal photons produced by the decays of the lightest (ordinary) neutrino is given by
\be
\varepsilon_{\g_{\rm nth}} = {\Delta m_1 \over \tau_1}\, n^{\infty}_{\nu_1}(z)\,\int_0^a \, da_D \, 
{{e^{-{t(a_D)}\over \tau_1}}\over H(a_D) a} \,   ,
\ee
where $a_D$ is the scale factor when neutrinos decay and $a$ is the scale factor at detection related to the redshift by $a = (1+z)^{-1}$. 
Notice that in the case of the excess radio background the detection is at the present time so that $z=0$ and $a = 1$. However, it will prove 
useful to be more general, considering cases when the detection occurred in the past (we will see in the next section how that is possible).  
Notice also that we denoted by $n^{\infty}_{\nu_1}(z)$ the standard  $\nu_1$ number density at redshift $z$ 
when neutrinos are stable given simply by
\be\label{nnu}
n^{\infty}_{\nu_1}(z) = {6\over 11}{\zeta(3)\over \pi^2}\,T^3(z) \,  .
\ee
From the expression for the energy density, it is easy to derive the following expression for the
specific intensity \cite{Masso:1999wj,Chianese:2018luo,Dev:2023wel}:
\be\label{Inth}
I_{\gamma_{\rm nth}}(E,z)  = {1 \over 4 \pi} \, {d\varepsilon_{\g_{\rm nth}} \over d E}
\, = {n_{\nu_1}^{\infty}(z) \over 4\,\pi} 
{e^{-{t(a_{\rm D}) \over \tau_i}} \over H(a_{\rm D}) \, \tau_1} \,  .
\ee
For $E \ll T_{\g_{\rm nth}}$, one obtains a simple linear relation between effective temperature and specific intensity
\be
T_{\g_{\rm nth}} \simeq {4\pi^3 \over E^2} \, I_{\g_{\rm nth}}(E,z) \,  .
\ee
In this more general case, the energy of the photon produced at redshift $z_{\rm D}$ and detected at redshift $z$ is given by 
\be
E(z) = \Delta m_1  {1 + z \over 1 + z_{\rm D}} \leq \Delta m_1 \,   .
\ee
For $z =0$, one recovers Eq.~(\ref{Eg0}).
The expansion rate at the decay, $H(a_{\rm D})$, can be calculated in the $\L$CDM model  as
\be\label{expr}
H(a_{\rm D}) = H_0\,\sqrt{\Omega_{{\rm M}0} \,a_{\rm D}^{-3} + \Omega_{\Lambda  0}} = 
H_0 \, \sqrt{\Omega_{M0}} \, a_{\rm D}^{-{3\over 2}} \, 
\left(1 + {a_{\rm D}^3 \over a_{\rm eq}^3} \right)^{{1 \over 2}}  \,  ,
\ee
where $a_{\rm eq} \equiv (\Omega_{M0}/\Omega_{\Lambda 0})^{1/3} \simeq 0.77$, 
$\Omega_{M0} \simeq 0.3111$, $H_0 \simeq t_0^{-1}$ and $t_0 \simeq 13.8 \, {\rm Gyr} \simeq 4.35 \times 10^{17}\,{\rm s}$ \cite{Planck:2018vyg}. One can also obtain
an analytical expression for the age of the universe at the time of decay, $t(a_{\rm D})$ \cite{DiBari:2018vba}:
\be
t(a_{\rm D}) = {2\over 3}\,{H_0^{-1}\over \sqrt{\Omega_{\Lambda 0}}}\,
\ln\left[\sqrt{\left({a_{\rm D}\over a_{\rm eq}}\right)^3} + \sqrt{1+\left({a_{\rm D}\over a_{\rm eq}}\right)^3} \right] \,  .
\ee
In order to fit the ARCADE 2 data, these expressions have to be specialised for $z=0$. The specific intensity, from Eq.~(\ref{Inth}), then becomes
\be\label{Inth}
I_{\gamma_{\rm nth}}(E,0)  =  {n_{\nu_1}^{\infty}(z) \over 4\,\pi} 
{e^{-{t(a_{\rm D}) \over \tau_i}} \over H(a_{\rm D}) \, \tau_1} \,  .
\ee
\begin{figure}
\begin{center}
 \psfig{file=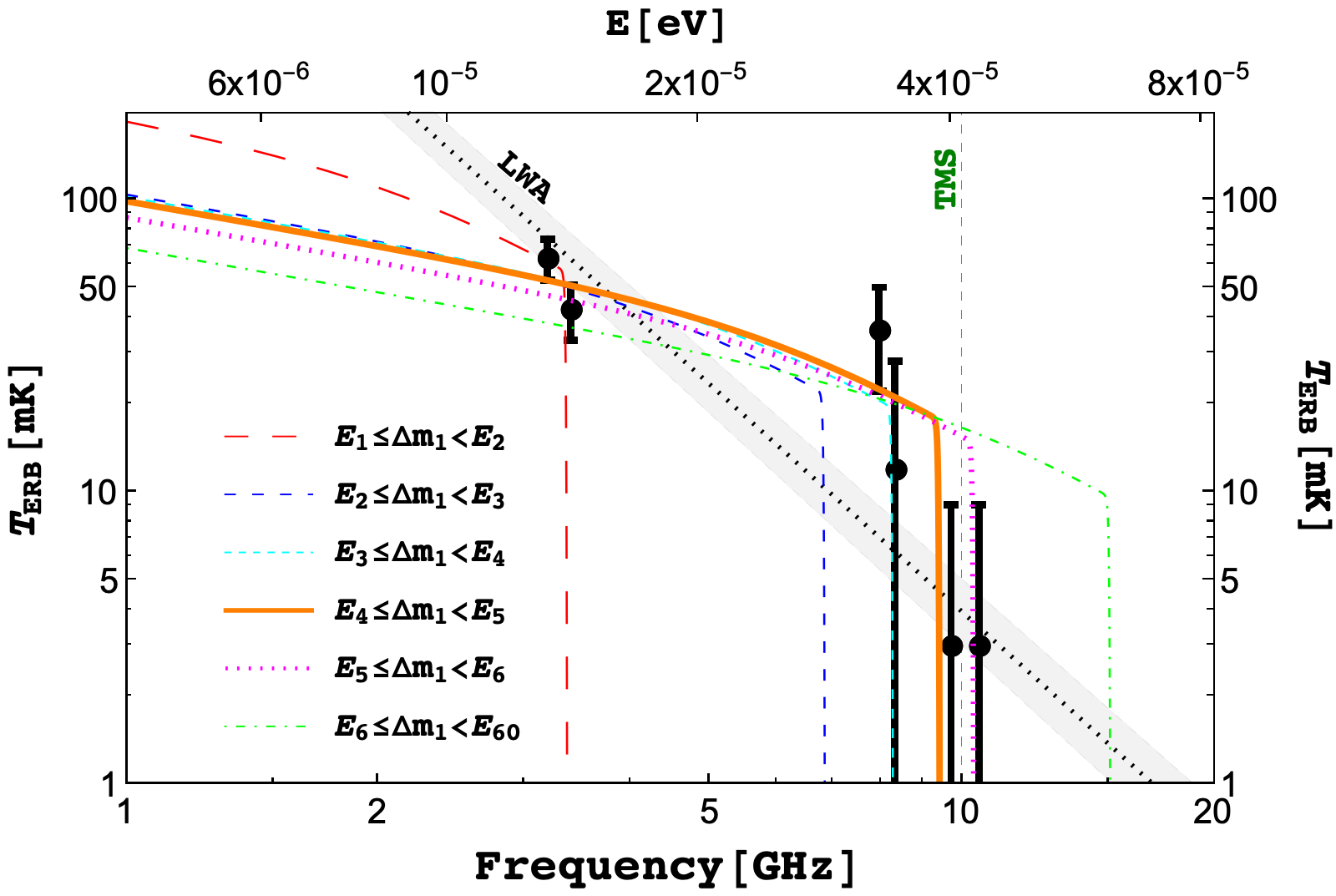,scale=0.6,angle=0}
\end{center}
    \caption{Best fit curves for $T_{\rm ERB}$ obtained with Eq.~(\ref{Tgammanth}). The thick solid orange curve corresponds to a solution very close to the
    best global fit ($\D m_1 = 4.0 \times 10^{-5}\,{\rm eV}$  and $ \tau_1= 1.46 \times 10^{21}\,{\rm s}$).
    The ARCADE 2 data points are taken from Ref.~\cite{Fixsen:2009xn}, while the power-law fit 
    $\beta = -2.58 \pm 0.05$ (dotted line with grey shade) is from \cite{Dowell:2018mdb}. The figure is taken from \cite{Dev:2023wel}.}
    \label{fig:bestfit}
\end{figure}
Assuming that the solution is found for $\tau_1 \gg \t_0 \geq t_{\rm D}$, we can also approximate the exponential to unity.
Using Eq.~(\ref{expr}) for the expansion rate at the decay time, we finally find for the effective temperature 
\be\label{Tgammanth}
T_{\gamma_{\rm nth}}(E,0) \simeq  {6\,\zeta(3)\over 11 \, \sqrt{\Omega_{{\rm M}0}}}\,
{T_{0}^3 \over E^{1 / 2}\, \Delta m_1^{3 / 2} } \, {t_0 \over \tau_1} 
\,\left(1 +{a_{\rm D}^3 \over a_{\rm eq}^3} \right)^{-{1 \over 2}} \,  .
\ee
When this is used to fit the six data points found by ARCADE 2, one finds the following best fit  \cite{Dev:2023wel}:
\be\label{bestfittau}
\tau_1 = 1.46  \times 10^{21}\,{\rm s}  \,  , \;\;\;\;  m_1 -m_{\rm s} = 4.0 \times 10^{-5}\,{\rm eV}
\ee
with a very good $\chi_{\rm min}^2/4{\rm d.of.} = 0.96$.  
The curve for the effective temperature for these best fit parameters  is shown in Fig.~3
with a orange thick solid line.  In the vertical axis $T_{\rm ERB}(E) \equiv T_{\gamma 0}(E) - T_{{\rm CMB},0}$ is the effective 
temperature of the excess radio background. It can be noticed how one of the most clear features of the fit is the existence of an
end-point at $E = m_1 -m_{\rm s}$.
For comparison, in Fig.~3 one one can also see the best fit in terms of a power law \cite{Dowell:2018mdb} with $\chi_{\rm min}^2/4{\rm d.of.} = 2.5$:
currently, the relic neutrino decay solution provides the best fit to the ARCADE 2 data.

{\em A clash with the upper bound on the effective magnetic moment}. However, the relic neutrino decay solution to the 
excess radio background mystery faces a challenging problem. Independently of the specific model one can build for neutrino radiative decays, 
one has to consider a general relation between decay rate $\Gamma_{\nu_i \ra \nu_j + \gamma}$ and effective neutrino magnetic moment 
$\mu_{{\rm eff}}^{i\!j}$
given by \cite{Mohapatra:1998rq,Xing:2011zza}
\be\label{gammamu}
\Gamma_{\nu_j \rightarrow \nu_i + \gamma} =  { \mu^{i\! j}_{\rm eff}\over 8\pi}\, \left({m^2_j - m^2_i \over m_j}\right)^3  \,  .
\ee
This relation implies that the upper bounds
\be\label{upperbmu}
\mu^{ij}_{\rm eff}\lesssim 3.2 \times 10^{-11}\,\mu_{\rm B} \, , \;\;\;\; \mu^{ij}_{\rm eff}\lesssim 3 \times 10^{-12}\,\mu_{\rm B}
\ee
placed, respectively, by neutrino-electron scattering experiments \cite{Beda:2010hk} and from plasmon decays in globular 
cluster stars \cite{Raffelt:1990pj}, translate into stringent lower bounds on 
the lifetime $\tau(\nu_2 \rightarrow \nu_1+\gamma) \equiv \Gamma^{-1}(\nu_2 \rightarrow \nu_1+\gamma)$ shown in Fig.~4.
How can this very challenging clash be solved?
\begin{figure}
\begin{center}
 \psfig{file=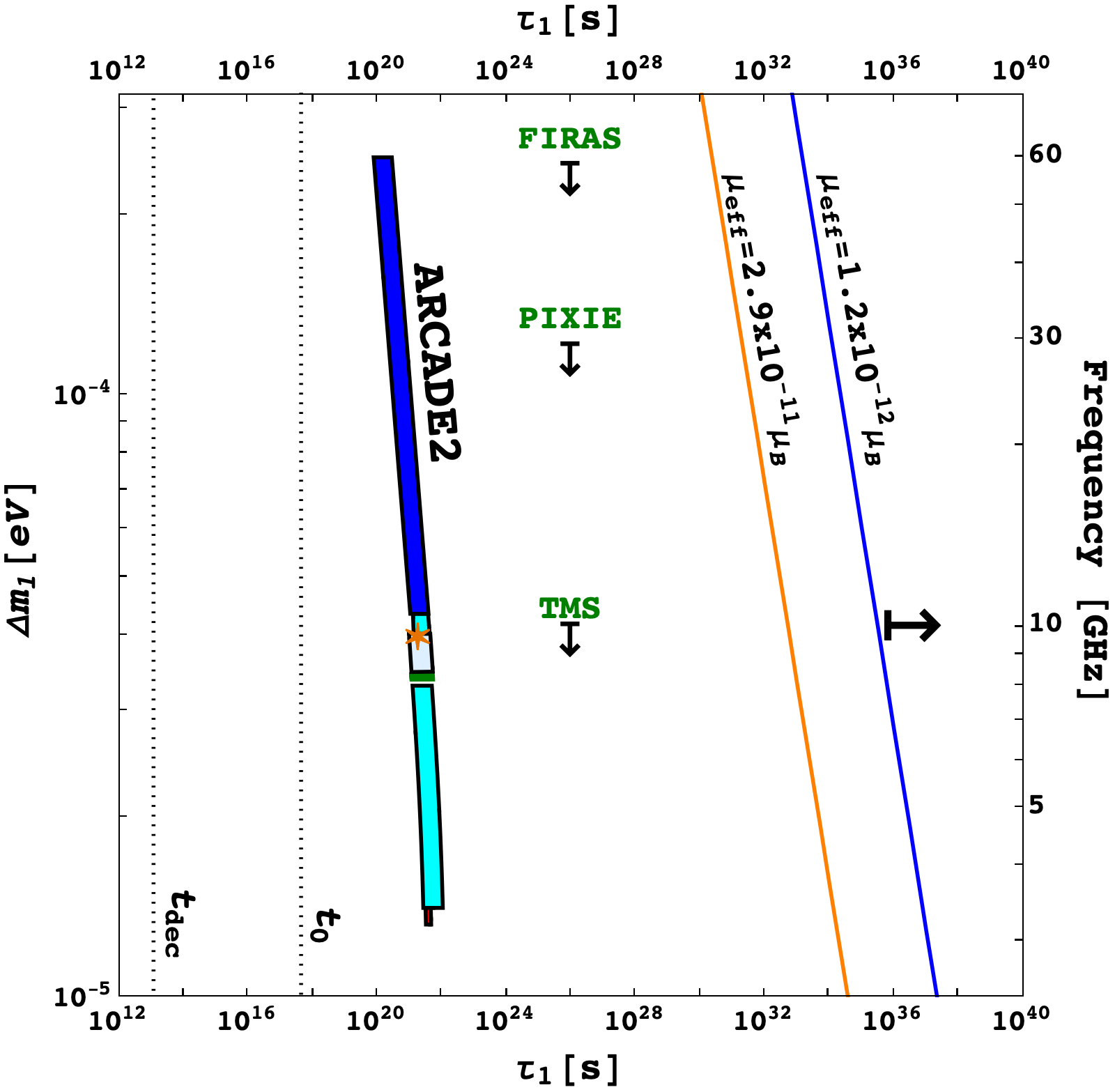,scale=0.45,angle=0}
\end{center}
    \caption{Lower bounds on the radiative neutrino decay lifetime derived from Eq.~(\ref{gammamu}) using the
    upper bounds in Eq.~(\ref{upperbmu}) on the effective neutrino magnetic moment.
    The figure is taken from \cite{Dev:2023wel}.}
    \label{fig:mueff}
\end{figure}

{\em Boomerang mechanism}. In order to circumvent the stringent constraint coming from Eq.~(\ref{gammamu}) combined with the
 upper bounds in Eq.~(\ref{upperbmu}) on the effective neutrino magnetic moment, the  
 boomerang mechanism sketched in Fig.~5 was  recently proposed \cite{Dev:2023wel}.  
\begin{figure}
\begin{center}
 \psfig{file=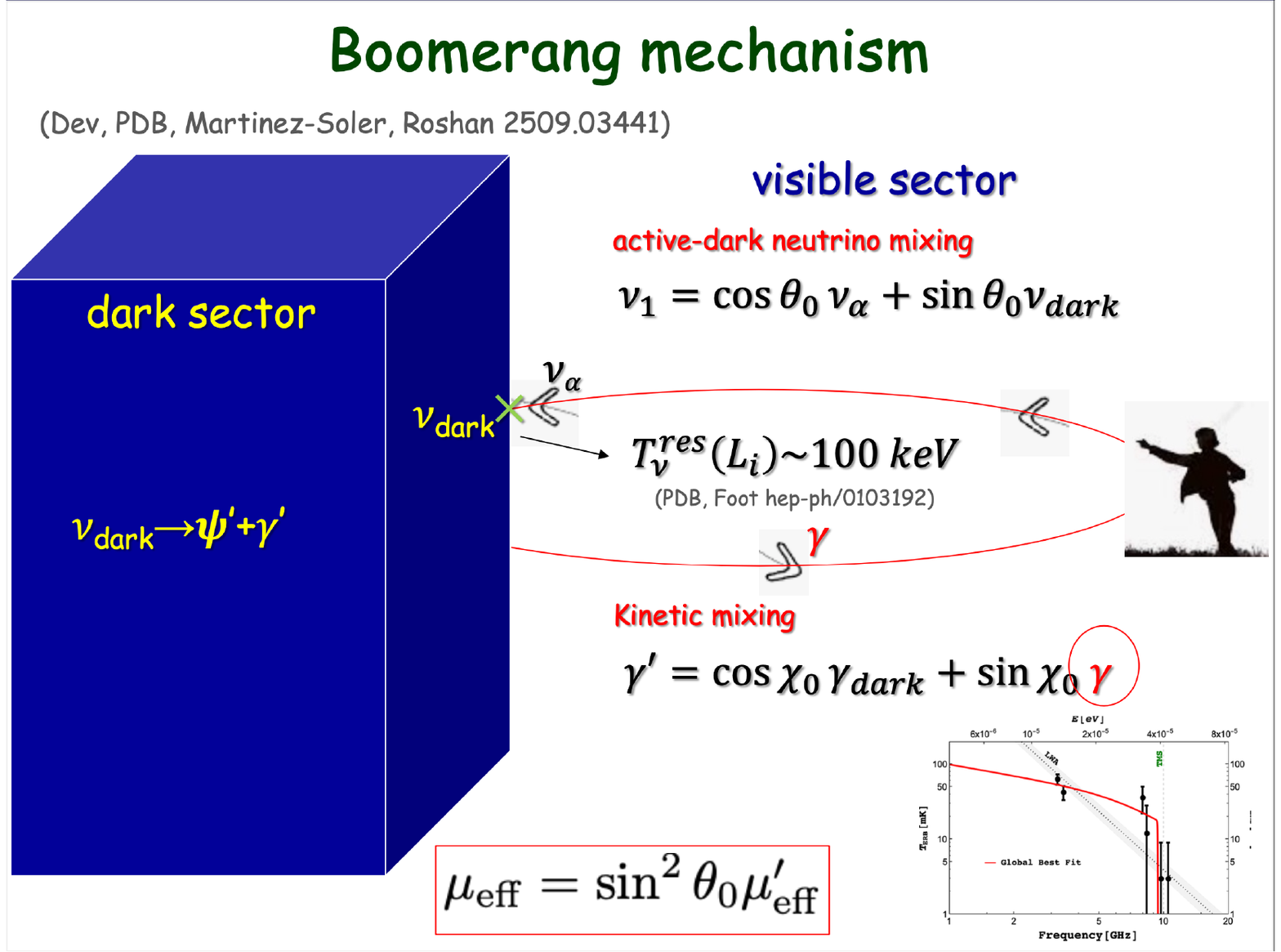,scale=0.35,angle=-90}
\end{center}
    \caption{Sketch of the boomerang mechanism proposed in \cite{Dev:2023wel}.}
    \label{fig:mueff}
\end{figure}
It is assumed that the active neutrino $\nu_\alpha$ ($\alpha = e,\mu,\tau$) is mixed with a dark neutrino $\nu_{\rm dark}$ with 
a small mixing angle $\theta_0$. The lightest neutrino mass eigenstate is then given by a linear combination
\be
\nu_1 = \cos\theta_0 \nu_{\alpha} + \sin\theta_0 \nu_{\rm dark} \,  .
\ee
In the presence of a large (effective) lepton asymmetry $L_{\rm i} \sim 10^{-4}$ and $\Delta m^2 = m^2_1 - m^2_0 \sim 10^{-4}\,{\rm eV^2}$, 
medium effects yield a resonance in the
early universe at $T \sim 100\,{\rm keV}$ \cite{Enqvist:1990ad}. If the mixing angle $\theta_0 \gtrsim 10^{-8}$, the resonance
is crossed adiabatically \cite{DiBari:2001jk} and at the resonance all $\nu_\alpha$ are converted into dark neutrinos $\nu_{\rm dark}$.
Dark neutrinos would then radiatively decay, with a value of the lifetime necessary to explain the excess radio background as in Eq.~(\ref{bestfittau}),  
into a quasi-degenerate dark fermion $\psi'$.  In this way the upper bound (\ref{upperbmu}) would not apply to the effective magnetic moment
of the active neutrino but to  that one of the dark neutrino, that we can denote by
$\mu'_{\rm eff}$, on which thee are not direct experimental constraints.
However, because of the active-dark neutrino mixing, the active neutrino still has an effective neutrino magnetic moment
$\mu_{\rm eff} = \sqrt{\ve}\sin^2\theta_0 \mu'_{\rm eff}$ and for this reason the experimental constraints on $\mu_{\rm eff}$ 
still play a role.  The name `boomerang' mechanism is justified by the fact that the visible sector first throws sterile neutrinos 
into the dark sector via mixing at a very early time when $T\sim 100\,{\rm keV}$ corresponding to $t \simeq 100\,{\rm s}$, 
 and then these, decaying, throw back photons into the visible sector. The net result is that neutrinos are converted into photons.
\footnote{Trading off the effective magnetic moment of ordinary neutrinos with that of sterile neutrino to circumvent (\ref{upperbmu})
 is an ingredient, the only one, in common with the model proposed in Ref.~\cite{AristizabalSierra:2018emu} to explain the EDGES anomaly.}
 In Fig.~6 one can see the constraints in plane of $\Delta m^2$ versus $\sin^2 2\theta_0$ (left) and
 $\mu_{\rm eff}$ versus $T_{\nu}^{\rm res}$, where we assumed $\alpha = \mu$ for definiteness.
\begin{figure*}[t!]
\begin{center}
  \includegraphics[scale=0.29]{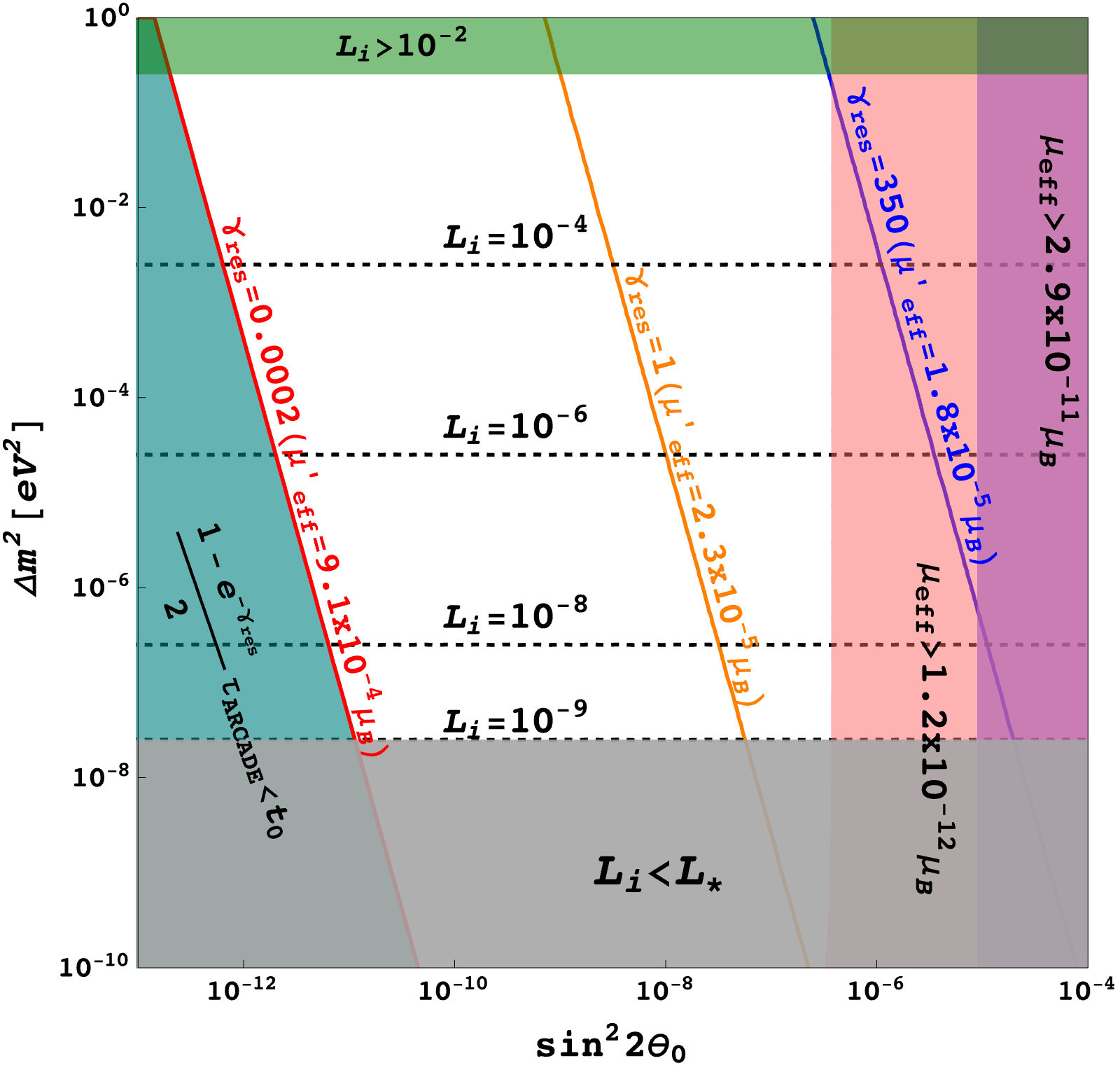}
   \includegraphics[scale=0.3]{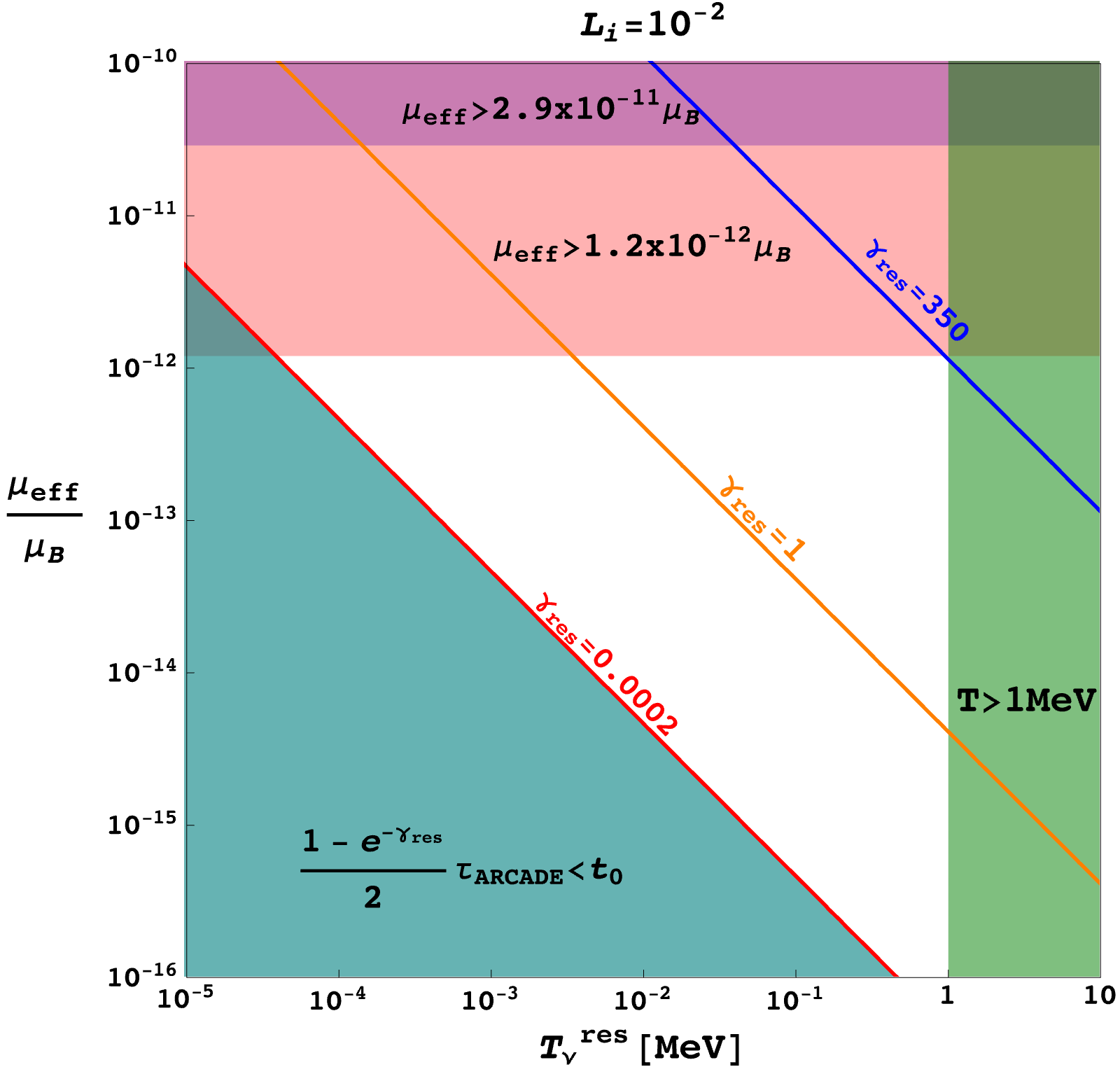}
\end{center}
    \caption{Left panel: Constraints (shaded) and allowed region (white) in the plane of $\Delta m^2$ versus $\sin^2 2\theta_0$. 
    The horizontal dashed lines give an upper bound on $\Delta m^2$ from $T_{\nu}^{\rm res}< 1\,{\rm MeV}$ for the indicated values of 
    the initial lepton asymmetry $L_{\rm i}$.
    Right panel: Constraints (shaded) and allowed region (white) in the plane of $\mu_{\rm eff}$ versus $T_{\nu}^{\rm res}$.
    The figure is taken from \cite{Dev:2023wel}.}
    \label{fig:constDm2sin2th0}
\end{figure*}
The assumption of a large initial lepton asymmetry could be regarded as an additional {\em ad hoc} ingredient of the mechanism.
However, it is quite remarkable that the same active-dark neutrino mixing, with the same mixing parameters, can
be responsible for the generation of a large lepton asymmetry \cite{Foot:1995qk,DiBari:2001jk}.

\section{21 cm cosmology: shedding light on dark ages}

The radiative decays of relic neutrinos would also leave an imprint in the 21 cm cosmological global signal \cite{Chianese:2018luo}. Indeed
the 21 cm cosmological global signal represents an important diagnostic tool to test new physics after the recombination era \cite{Pritchard:2011xb}.
The 21 cm (emission or absorption) line is produced by hyperfine transitions between the 
spin-singlet and triplet energy levels of the 1s ground state of hydrogen atoms. 
The energy splitting between the two levels is $E_{21} = 5.87\, \mu{\rm eV}$, corresponding 
to a rest frequency $\nu_{21}^{\rm rest}= 1.420\, {\rm GHz}$. 
This transition can be used cosmologically to obtain information on the cosmological history and parameters in a
wide redshift range $z \sim 7$--$200$. 
The 21 cm brightness temperature parameterises the brightness contrast 
between the cosmic radiation and the absorbed or emitted radiation in the 21 cm transitions 
is usually referred to as the 21 cm cosmological global signal and is approximately given by \cite{Zaldarriaga:2003du}
\be\label{T21}
T_{21}(z) \simeq  23\,{\rm mK} \, (1+\d_{\rm B})\, x_{H_I}(z) \,\left({\Omega_{{\rm B}0} h^2 \over 0.02}\right)\,
\left[\left({0.15 \over \Omega_{{\rm M}0}h^2}\right)\,
\left({1+z \over 10} \right)  \right]^{1/2}  \,\left[
1 - {T_\g(z) \over T_{\rm S}(z)} \right]  \,  ,
\ee
where $T_{\rm S}(z)$ is the spin temperature describing the  triplet-to-singlet state density ratio. 
In this expression $\d_{\rm B} = (\rho_{\rm B}-\bar{\rho}_{\rm B})/\bar{\rho}_{\rm B}$ is the fractional
baryon overdensity, $x_{H_I}(z)$ is the neutral hydrogen fraction.
If the spin temperature is equal to the photon temperature ($T_{\rm S} = T_{\gamma}$), 
then photons are absorbed and reemitted with the same intensity and there is no visible signal. 
Also, if all atoms are ionised so that $x_{H_I} = 0$, there cannot be any signal. 

The most prominent feature that is expected within the $\Lambda$CDM model in the 21 cm cosmological signal is an absorption 
signal, corresponding to a negative value of  $T_{21}$, at redshifts in the range $z=10$--$30$. 
The $\Lambda$CDM model predicts a relic photon temperature
\be
T_{\gamma}(\bar{z}_E) = T(z_E) = T_0 \, (1+z_E) \simeq 49.6 \, {\rm K} \,  ,
\ee
and a gas temperature $T_{\rm gas}(\bar{z}_E) \simeq 7.2 \, {\rm K}$, where $z_E \simeq 17$. 
Defining  $\xi(z) \equiv T_{\gamma}(z)/T_{\rm gas}(z)$, one has then $\xi(\bar{z}_E) \simeq 6.89$, 
corresponding, to $T_{21}(\bar{z}_E) \simeq -206\,{\rm mK}$ from Eq.~(\ref{T21}). The EDGES collaboration  
claimed to have detected such absorption feature  centred at $z = z_{\rm E} \simeq 17$ \cite{EDGES}, 
thus within the expected range of redshifts. However, the measured (negative) value of $T_{21}$ is approximately
twice the expected one:
\be
T_{21}^{\rm EDGES}(z_E) = -500^{+200}_{-500}\,{\rm mK} \hspace{10mm} (99\%\,{\rm C.L.})  \,  .
\ee 
This anomalous signal can be interpreted in terms of a non-thermal radiation component produced by relic neutrino decays with 
a temperature, at $z =z_E$, $T_{\gamma_{\rm nth}} \simeq 60\,{\rm K}$ \cite{Chianese:2018luo}. 
If we continue to assume, as in the case of the solution to the excess radiobackground mystery, 
that the decaying neutrinos are the lightest active neutrinos, decaying 
non-relativistically into quasi-degenerate sterile neutrinos, then
Eq.~(\ref{Inth}) has to be specialised to the case $z = z_{\rm E}$ and $E = E_{21} = 5.87\mu {\rm eV}$ and imposing
$T_{\gamma_{\rm nth}}(z=z_E) \simeq 60\,{\rm K}$, one obtains 
\be\label{DmtauEDGES}
(\D m_1^{3/2}\tau_1)^{\rm EDGES} \simeq 4.0 \times 10^{13}\,\,{\rm eV}^{3/2}\,{\rm s}  \,  .
\ee
This should be compared with the value for the same quantity that is obtained as best fit of the ARCADE 2 data \cite{Dev:2023wel}:
\be\label{ARCADE}
(\D m_1^{3/2}\tau_1)^{\rm ARCADE 2}  = 3.8^{+7.2}_{-1.5}  \times 10^{14}\,{\rm eV}^{3/2}\,{\rm s} \,  .
\ee
It is evident that there is a  tension between the two results. This can be also expressed deriving the value 
$T_{21}$ at $z_E$ predicted by the solution to the ARCADE 2 data:
\be\label{T21predicted}
T_{21}^{\rm ARCADE 2}(\bar{z}_E) = -238^{+21}_{-20}\,{\rm mK} \;\;\; (99\%\,{\rm C.L.}) . 
\ee  
This value represents just a small deviation from the $\L$CDM prediction, in contrast to the large deviation requested by the EDGES anomaly.
At face value the ARCADE 2 and the EDGES anomalies do not seem compatible within the relic neutrino decay solution. However,
the EDGES anomaly is controversial and a few studies have suggested that the signal is contaminated by some foreground contribution,
for example originating in the ionosphere.  The SARAS3 experiment has even rebutted the EDGES claim \cite{Singh:2021mxo}, so we need to wait for 
more results. In this respect it is exciting that the moon-based experiment LuSEE will soon be able to measure the 21 cm cosmological global signal in absence of
Earth foregrounds  \cite{2023arXiv230110345B}.

\section{Conclusions}

Let us draw some final remarks:
\begin{itemize}
\item Neutrino can help solving or alleviating different anomalies and tensions in cosmology.
\item Invisible decays of the heaviest relic neutrinos might provide a way to solve the
neutrino mass tension between cosmological observations and neutrino oscillation experiments.
\item Also the excess radio background mystery could be explained by radiative decays of relic neutrinos.
However, the upper bound on the neutrino effective magnetic moment requires some trick to be circumvented.
\item We proposed a boomerang mechanism: the visible sector throws dark neutrinos into the dark sector at $t \sim 100\,{\rm s}$
and $T \sim 100\,{\rm keV}$, and much later (basically at the present time) the dark sector throws back photons into the visible sector.  
\item The mechanism predicts an effective neutrino magnetic moment that might be within the reach of next experiments.
\item Some contribution to the 21 cm cosmological signal is also expected, though to a much lower level than claimed by EDGES.
\end{itemize}
In conclusion: these are exciting times for cosmological searches of BSM physics. 

\subsubsection*{Acknowledgments}
I acknowledge financial support from the STFC Consolidated Grant ST/T000775/1.
I also acknowledge support from the European Union’s Horizon 2020 Europe research and innovation programme under  
the Marie Sk\l odowska-Curie grant agreement HIDDeN European  ITN project (H2020-MSCA-ITN2019//860881-HIDDeN). 
It is a pleasure to thank Marco Chianese, Bhupal Dev, Kareem Farrag, Ivan Mart\'\i{}nez-Soler,
Rishav Roshan and Rome Samanta for a fruitful collaboration on the topics discussed in the talk.


\begin{thebibliography}{99}

\bibitem{Dev:2023wel}
P.~S.~B.~Dev, P.~Di Bari, I.~Mart\'\i{}nez-Soler and R.~Roshan,
{\em Relic neutrino decay solution to the excess radio background},
JCAP \textbf{04} (2024), 046
[arXiv:2312.03082 [hep-ph]].

\bibitem{Dev:2025ufo}
B.~Dev, P.~Di Bari, I.~Martinez-Soler and R.~Roshan,
{\em Boomerang mechanism explaining the excess radio background},
to appear in PRD, [arXiv:2509.03441 [hep-ph]].

\bibitem{Esteban:2024eli}
I.~Esteban, M.~C.~Gonzalez-Garcia, M.~Maltoni, I.~Martinez-Soler, J.~P.~Pinheiro and T.~Schwetz,
{\em NuFit-6.0: updated global analysis of three-flavor neutrino oscillations},
JHEP \textbf{12} (2024), 216
[arXiv:2410.05380 [hep-ph]].


\bibitem{Tristram:2023haj}
M.~Tristram, A.~J.~Banday, M.~Douspis, X.~Garrido, K.~M.~G{\'o}rski, S.~Henrot-Versill{\'e}, L.~T.~Hergt, S.~Ili{\'c}, R.~Keskitalo and G.~Lagache, \textit{et al.}
{\em Cosmological parameters derived from the final Planck data release (PR4)},
Astron. Astrophys. \textbf{682} (2024), A37
[arXiv:2309.10034 [astro-ph.CO]].

\bibitem{DESI:2025zgx}
M.~Abdul Karim \textit{et al.} [DESI],
{\em DESI DR2 results. II. Measurements of baryon acoustic oscillations and cosmological constraints},
Phys. Rev. D \textbf{112} (2025) no.8, 083515
[arXiv:2503.14738 [astro-ph.CO]].

\bibitem{Giare:2025ath}
W.~Giar{\`e}, O.~Mena, E.~Specogna and E.~Di Valentino,
{\em Neutrino mass tension or suppressed growth rate of matter perturbations?},
Phys. Rev. D \textbf{112} (2025) no.10, 103520
[arXiv:2507.01848 [astro-ph.CO]].

\bibitem{Hannestad:2005ex}
S.~Hannestad and G.~Raffelt,
{\em Constraining invisible neutrino decays with the cosmic microwave background},
Phys. Rev. D \textbf{72} (2005), 103514
[arXiv:hep-ph/0509278 [hep-ph]].

\bibitem{Serpico:2007pt}
P.~D.~Serpico, {\em Cosmological Neutrino Mass Detection: The Best Probe of Neutrino Lifetime},
Phys. Rev. Lett. \textbf{98} (2007), 171301
[arXiv:astro-ph/0701699 [astro-ph]].

\bibitem{Archidiacono:2013dua}
M.~Archidiacono and S.~Hannestad,
{\em Updated constraints on non-standard neutrino interactions from Planck},
JCAP \textbf{07} (2014), 046
[arXiv:1311.3873 [astro-ph.CO]].

\bibitem{Escudero:2019gfk}
M.~Escudero and M.~Fairbairn,
{\em Cosmological Constraints on Invisible Neutrino Decays Revisited},
Phys. Rev. D \textbf{100} (2019) no.10, 103531
[arXiv:1907.05425 [hep-ph]].


\bibitem{Escudero:2020ped}
M.~Escudero, J.~Lopez-Pavon, N.~Rius and S.~Sandner,
{\em Relaxing Cosmological Neutrino Mass Bounds with Unstable Neutrinos},
JHEP \textbf{12} (2020), 119
[arXiv:2007.04994 [hep-ph]].

\bibitem{Craig:2024tky}
N.~Craig, D.~Green, J.~Meyers and S.~Rajendran,
{\em No {\ensuremath{\nu}}s is Good News},
JHEP \textbf{09} (2024), 097
[arXiv:2405.00836 [astro-ph.CO]].

\bibitem{Fixsen:2009ug}
D.~J.~Fixsen,
{\em The Temperature of the Cosmic Microwave Background},
Astrophys. J. \textbf{707} (2009), 916-920
[arXiv:0911.1955 [astro-ph.CO]].

\bibitem{Fixsen:2009xn}
D.~J.~Fixsen, A.~Kogut, S.~Levin, M.~Limon, P.~Lubin, P.~Mirel, M.~Seiffert, J.~Singal, E.~Wollack and T.~Villela, \textit{et al.}
{\em ARCADE 2 Measurement of the Extra-Galactic Sky Temperature at 3-90 GHz},
Astrophys. J. \textbf{734} (2011), 5
[arXiv:0901.0555 [astro-ph.CO]].

\bibitem{Grebenev:2024nkw}
S.~A.~Grebenev and R.~A.~Sunyaev,
{\em Increase in the Brightness of the Cosmic Radio Background toward Galaxy Clusters},
Astron. Lett. \textbf{50} (2024), 159-185
[arXiv:2408.01858 [astro-ph.HE]].

\bibitem{Chianese:2018luo}
M.~Chianese, P.~Di Bari, K.~Farrag and R.~Samanta,
{\em Probing relic neutrino radiative decays with 21 cm cosmology},
Phys. Lett. B \textbf{790} (2019), 64-70
[arXiv:1805.11717 [hep-ph]].

\bibitem{Masso:1999wj}
E.~Masso and R.~Toldra,
{\em Photon spectrum produced by the late decay of a cosmic neutrino background},
Phys. Rev. D \textbf{60} (1999), 083503
[arXiv:astro-ph/9903397 [astro-ph]].

\bibitem{Planck:2018vyg}
N.~Aghanim \textit{et al.} [Planck],
{\em Planck 2018 results. VI. Cosmological parameters},
Astron. Astrophys. \textbf{641} (2020), A6
[erratum: Astron. Astrophys. \textbf{652} (2021), C4]
[arXiv:1807.06209 [astro-ph.CO]].

\bibitem{DiBari:2018vba}
P.~Di Bari,
{\em Cosmology and the Early Universe},
CRC Press, 2018, ISBN 978-1-4987-6170-3, 978-1-138-49690-3.

\bibitem{Dowell:2018mdb}
J.~Dowell and G.~B.~Taylor,
{\em The Radio Background Below 100 MHz},
Astrophys. J. Lett. \textbf{858} (2018) no.1, L9
[arXiv:1804.08581 [astro-ph.CO]].

\bibitem{Mohapatra:1998rq}
R.~N.~Mohapatra and P.~B.~Pal,
{\em Massive neutrinos in physics and astrophysics. Second edition},
World Sci. Lect. Notes Phys. \textbf{60} (1998), 1-397


\bibitem{Xing:2011zza}
Z.~z.~Xing and S.~Zhou,
{\em Neutrinos in particle physics, astronomy and cosmology}.

\bibitem{Beda:2010hk}
A.~G.~Beda, V.~B.~Brudanin, V.~G.~Egorov, D.~V.~Medvedev, V.~S.~Pogosov, M.~V.~Shirchenko and A.~S.~Starostin,
{\em Upper limit on the neutrino magnetic moment from three years of data from the GEMMA spectrometer},
[arXiv:1005.2736 [hep-ex]].

\bibitem{Raffelt:1990pj}
G.~G.~Raffelt,
{\em New bound on neutrino dipole moments from globular cluster stars},
Phys. Rev. Lett. \textbf{64} (1990), 2856-2858

\bibitem{Enqvist:1990ad}
K.~Enqvist, K.~Kainulainen and J.~Maalampi,
{\em Refraction and Oscillations of Neutrinos in the Early Universe},
Nucl. Phys. B \textbf{349} (1991), 754-790

\bibitem{DiBari:2001jk}
P.~Di Bari and R.~Foot,
{\em Active sterile neutrino oscillations in the early universe: Asymmetry generation at low |delta m**2| and the Landau-Zener approximation},
Phys. Rev. D \textbf{65} (2002), 045003
[arXiv:hep-ph/0103192 [hep-ph]].

\bibitem{AristizabalSierra:2018emu}
D.~Aristizabal Sierra and C.~S.~Fong,
{\em The EDGES signal: An imprint from the mirror world?},
Phys. Lett. B \textbf{784} (2018), 130-136
[arXiv:1805.02685 [hep-ph]].

\bibitem{Foot:1995qk}
R.~Foot, M.~J.~Thomson and R.~R.~Volkas,
{\em Large neutrino asymmetries from neutrino oscillations},
Phys. Rev. D \textbf{53} (1996), R5349-R5353
[arXiv:hep-ph/9509327 [hep-ph]].

\bibitem{Pritchard:2011xb}
J.~R.~Pritchard and A.~Loeb,
{\em 21-cm cosmology},
Rept. Prog. Phys. \textbf{75} (2012), 086901
[arXiv:1109.6012 [astro-ph.CO]].

\bibitem{Zaldarriaga:2003du}
M.~Zaldarriaga, S.~R.~Furlanetto and L.~Hernquist,
{\em 21 Centimeter fluctuations from cosmic gas at high redshifts},
Astrophys. J. \textbf{608} (2004), 622-635
[arXiv:astro-ph/0311514 [astro-ph]].

\bibitem{EDGES}
J.~D.~Bowman, A.~E.~E.~Rogers, R.~A.~Monsalve, T.~J.~Mozdzen and N.~Mahesh,
 {\em An absorption profile centred at 78 megahertz in the sky-averaged spectrum},
  Nature {\bf 555} (2018) no.7694,  67.

\bibitem{Singh:2021mxo}
S.~Singh, J.~Nambissan T., R.~Subrahmanyan, N.~Udaya Shankar, B.~S.~Girish, A.~Raghunathan, 
R.~Somashekar, K.~S.~Srivani and M.~Sathyanarayana Rao,
{\em On the detection of a cosmic dawn signal in the radio background},
Nature Astron. \textbf{6} (2022) no.5, 607-617
[arXiv:2112.06778 [astro-ph.CO]].


\bibitem{2023arXiv230110345B}
S.D.~{Bale}, N.~{Bassett}, J.O.~{Burns}, J.~{Dorigo Jones}, K.~{Goetz},
  C.~{Hellum-Bye} et~al., \emph{{LuSEE 'Night': The Lunar Surface
  Electromagnetics Experiment}},
  \href{https://doi.org/10.48550/arXiv.2301.10345}{\emph{arXiv e-prints} (2023)
  arXiv:2301.10345} [\href{https://arxiv.org/abs/2301.10345}{{\ttfamily
  2301.10345}}].


\end{thebibliography}
\end{document}